\documentclass[journal=jpclcd,manuscript=article]{achemso}

\usepackage[version=3]{mhchem} 
\usepackage{graphicx}
\usepackage{dcolumn}
\usepackage{color}
\usepackage{hyperref}
\usepackage{mathtools}
\usepackage{array}
\usepackage{makecell}

\author{Azim Fitri Zainul Abidin}
\affiliation[Osaka University, Japan]
{Department of Precision Engineering, Graduate School of Engineering, Osaka University, 2-1 Yamadaoka, Suita, Osaka, 565-0871, Japan}

\author{Ikutaro Hamada}
\email{ihamada@prec.eng.osaka-u.ac.jp}
\phone{}
\fax{}
\affiliation[Osaka University, Japan]
{Department of Precision Engineering, Graduate School of Engineering, Osaka University, 2-1 Yamadaoka, Suita, Osaka, 565-0871, Japan}

\title{Oxygen Reduction Reaction on Single-Atom Catalysts From Density Functional Theory Calculations Combined with an Implicit Solvation Model}

\keywords{Oxygen reduction reaction, single atom catalyst, electrochemistry, density functional theory, implicit solvation model}

\begin{document}

\begin{tocentry}

\includegraphics[width=1.01\columnwidth]{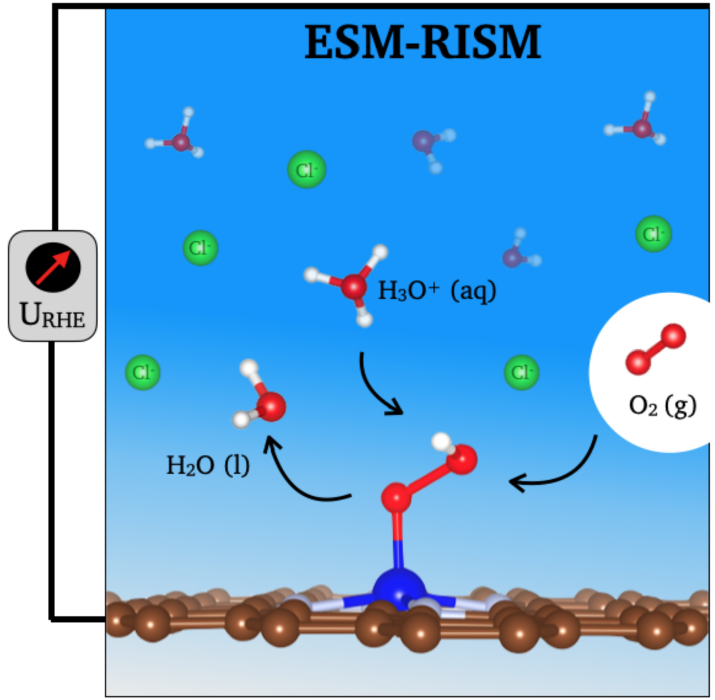}





\end{tocentry}


\begin{abstract}
We present a density functional theory study of the oxygen reduction reaction (ORR) on a single atom catalyst embedded in graphene, namely, TM$-$\ce{N4}$-$C (TM = Fe and Co), using the effective screening medium method combined with the reference interaction site model (ESM-RISM).
It was found that Fe$-$\ce{N4}$-$C and Co$-$\ce{N4}$-$C show comparable ORR activities from the constant electrode potential simulations, in contrast to the results obtained using the constant (neutral) charge simulation, in which the superior performance of Co$-$\ce{N4}$-$C has been predicted.
The constant potential method allows the variable charge and thus results in a potential dependence of the reaction-free energies different from that with the constant charge method in which the potential dependence is included as an ad hoc manner.
We suggest the importance of the variable charge in the simulation of the electrochemical reaction, which is enabled by ESM-RISM.
\end{abstract}
%

\newpage

\section{Introduction}

It has been well established that the single-atom catalyst, particularly the transition metal (TM = Fe and Co) embedded in N-doped graphene, has the potential to reduce \ce{O2} to \ce{H2O} with appreciable activity and stability\cite{lin2014noble,lefevre2009iron,xie2020performance,wu2011high}.
In the past decade, great research efforts have been dedicated to the development of this electrocatalyst, especially in searching and tuning the active center moieties of the TM$-$N$_x-$C$_y$ (where $x$ $\in$ [1, 2, 3, 4,..] and $y$ $\in$ [edge, 8, 10, 12,..]), as an effort to optimize their intrinsic activity for the proton exchange membrane fuel cell (PEMFC) application \cite{patniboon2021acid,jia2015experimental,chen2022identification,zitolo2015identification,zitolo2017identification}.
The state-of-the-art of this single-atom catalyst, mainly the 4-fold coordinate metal atom with pyridinic nitrogen atoms in graphene (Fe$-$N$_4-$C and Co$-$N$_4-$C), exhibits initial activity comparable to that of commercial Pt/C and offers an appealing pathway to solving the energy and environmental problem \cite{xie2020performance,mehmood2022high,abidin2019nitrogen}.
However, substantial improvements in the oxygen reduction reaction (ORR) activity and stability at lower energy costs, i.e., lower overpotentials, are still required to escalate the economic feasibility of the PEMFC system for going to scale \cite{wan2022iron,li2022identifying,fitri2017cobalt,abidin2018effect}.
The electrochemical ORR occurs at the electrode$-$electrolyte interface, and thus, both the electrode surface and electrolyte environment could play a vital role in the catalytic performance of the catalyst\cite{spendelow2007electrocatalysis,ramaswamy2011influence,tylus2014elucidating}.
Especially, the ORR activity on this catalyst has been shown to have a prominent sensitivity towards the changes in the environment (e.g., pH values, ionic effects, and so on)\cite{ramaswamy2013activity,bae2020ph,svane2019improving,zamora2021probing}.
It is believed that, under the ORR reducing potential, the catalytic reaction site will gain more charges and the electrical double-layer (EDL) is reconstructed due to the interface polarization, affecting the ORR intermediates at the electrochemical interface, and thus influencing the catalytic activity\cite{qian2022critical}.
However, it is challenging to experimentally characterize the detail of ORR under the working condition.
Therefore, atomistic simulations that take into account the electrochemical environment, i.e., electrode potential and EDL, are desirable to better understand the underlying reaction mechanism of ORR.

The computational hydrogen electrode (CHE) model\cite{norskov2004origin} is arguably the most widely used atomistic approach to studying the mechanisms of electrochemical reactions and exploring efficient electrocatalysts.
Within the conventional CHE model, the catalyst and reaction intermediates are treated quantum mechanically (mostly) in the vacuum condition, and the effects of the applied potential and electrolyte concentration (pH value) are included posterior as an algebraic correction to the free energy.
The solvent effect can be included in the CHE model through an implicit solvent model or explicit solvent through (either classical or quantum mechanical) atomistic simulations, but the latter can be computationally extensive.
There have been several methodological developments to include the effects of the electrode potential\cite{lozovoi2001ab,otani2006first,bonnet2012first,sundararaman2017grand} and the electrolyte solution\cite{otani2006first,taylor2006first,otani2008electrode,jinnouchi2008electronic,letchworth2012joint,nishihara2017hybrid,melander2019grand,hormann2019grand} in simulations of electrode$-$electrolyte solution interfaces within the grand-canonical ensemble for electrons and/or ions.
In particular, the progress of the hybrid quantum-mechanical and molecular-mechanical method, in which electrode surface and adsorbate(s) are treated quantum mechanically, while the electrolyte solution is described using continuum model or solution theory\cite{letchworth2012joint,sundararaman2017grand,nishihara2017hybrid,melander2019grand,hormann2019grand}, is remarkable.
In this study, we investigate the mechanism of ORR on TM$-$\ce{N4}$-$C (TM = Fe, Co) catalysts by means of the density functional theory (DFT) calculations combined with the effective screening medium (ESM) method and the reference interaction site model (RISM),\cite{nishihara2017hybrid} a hybrid DFT and an implicit solvation model.
By employing the fictitious charge particle method developed by Bonnet et al.\cite{bonnet2012first}, a constant-potential (constant-$\mu_{e}$) simulation (i.e., variable number of electrons/surface charge) is made possible.
Furthermore, RISM allows for a variable number of electrolyte ions in solution in response to the surface charge, thus keeping the charge neutrality of the whole system and the electrode potential of the electrode$-$electrolyte solution system can be defined.\cite{haruyama2018electrode}
%

%
%
\section{Computational method}
We used the projector augmented wave (PAW) method\cite{blochl1994projector} as implemented in \textsc{Quantum-ESPRESSO} code\cite{giannozzi2009quantum,giannozzi2017advanced}.
PAW potentials were adopted from \textsc{pslibrary}\cite{dal2014pseudopotentials}, which were generated using the Perdew$-$Burke$-$Ernzheof (PBE)\cite{perdew1996generalized} generalized gradient approximation to the exchange-correlation (XC) functional.
Wave functions and augmentation charge density were expanded in terms of a plane-wave basis set with the kinetic energy cutoffs of 80 Ry and 800 Ry, respectively. 
The Fermi surface was treated using the Marzari$-$Vanderbilt cold smearing with a smearing width of 0.02 Ry\cite{marzari1999thermal}.
The revised PBE exchange-correlation functional of Hammer et al.\cite{hammer1999improved} with the dispersion correction of Grimme\cite{grimme2010consistent} (RPBE-D3), which was proven\cite{abidin2022comparative} to show comparative accuracy with the BEEF-vdW functional\cite{wellendorff2012density}.
We confirmed that the use of PBE potentials does not change the results significantly in the RPBE(-D3) calculations.\cite{abidin2022comparative}
%
The TM$-$\ce{N4} moiety and reaction intermediates were described by DFT using a graphene (5 $\times$ 5) supercell, which was in contact with the HCl solution at 298.15~K described by RISM.
The kinetic energy cutoff for the solvent correlation functions was set to 160 Ry.
We used the vacuum/slab/solvent boundary condition to model TM$-$\ce{N4}$-$C's and reaction intermediates, while solvated 
\ce{H2O} and \ce{H3O+} were modeled with the solvent/ion/solvent one.\cite{haruyama2018electrode,kano2021study}
The electrolyte solution was described with simple point charge and five-point intermolecular potential models with modified Lennard-Jones (LJ) parameters for \ce{H2O}\cite{haruyama2018electrode}
and those for \ce{H3O+} and \ce{Cl-} adopted from Chuev et al.\cite{chuev2006quasilinear} and Smith and Dang.\cite{smith1994computer}.
%
%
The LJ parameters for C and N atoms were generated using the data from Ref.~\cite{abidin2022interaction}
, and those for Fe and Co atoms were generated using the interaction energy curves for a \ce{H2O} on Fe$-\ce{N4}-$C and Co$-\ce{N4}-$C surfaces in the most stable configurations.
See Table S1 for the LJ parameters generated and used in this study.
%
%
We chose to use the HCl solution to represent an acidic environment. Note that the electrolyte plays an essential role in determining the electrode potential and pH, but the poisoning effect of the anion reported in Refs.~\cite{holst2018enhanced,patniboon2023effects} was not considered in the modeling with plain RISM employed in the present study.
%
%
A 4$\times$4 $\mathbf{k}$-point grid was used to sample the surface Brillouin zone.
Structural optimization was performed until the residual forces became less than $2.57 \times 10^{-2}$ eV/\AA~($10^{-3}$ Ry/$a_0$).
See the Supporting Information for further details of the calculation.
%
%
\section{Results and discussion}
\subsection{Electric double layer}
\begin{figure}[h]
   \includegraphics[width=1\columnwidth]{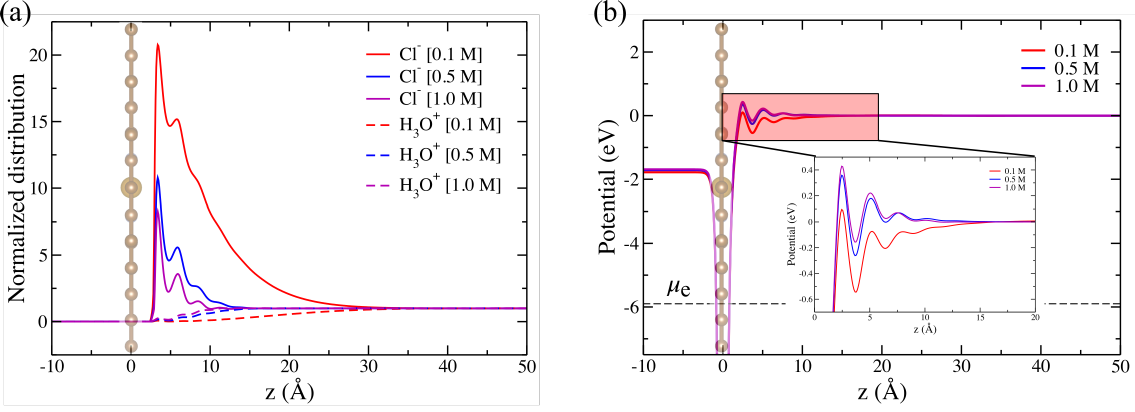}
   \caption{\label{fig:distribution} (a) Distribution of \ce{H3O+} and \ce{Cl-} in  HCl solution at 0.7 V vs RHE. (b) Electrostatic potential at 0.7 V vs RHE of Fe$-$\ce{N4}$-$C in contact with the HCl solution of the electrolyte concentrations of 1.0, 0.5, and 0.1 M. The Fermi energy ($\mu_{\mathrm{e}}$) is depicted by the horizontal dashed line. The graphene layer is depicted by a ball-and-stick model.} 
\end{figure}

\begin{figure}[ht]
   \includegraphics[width=1\columnwidth]{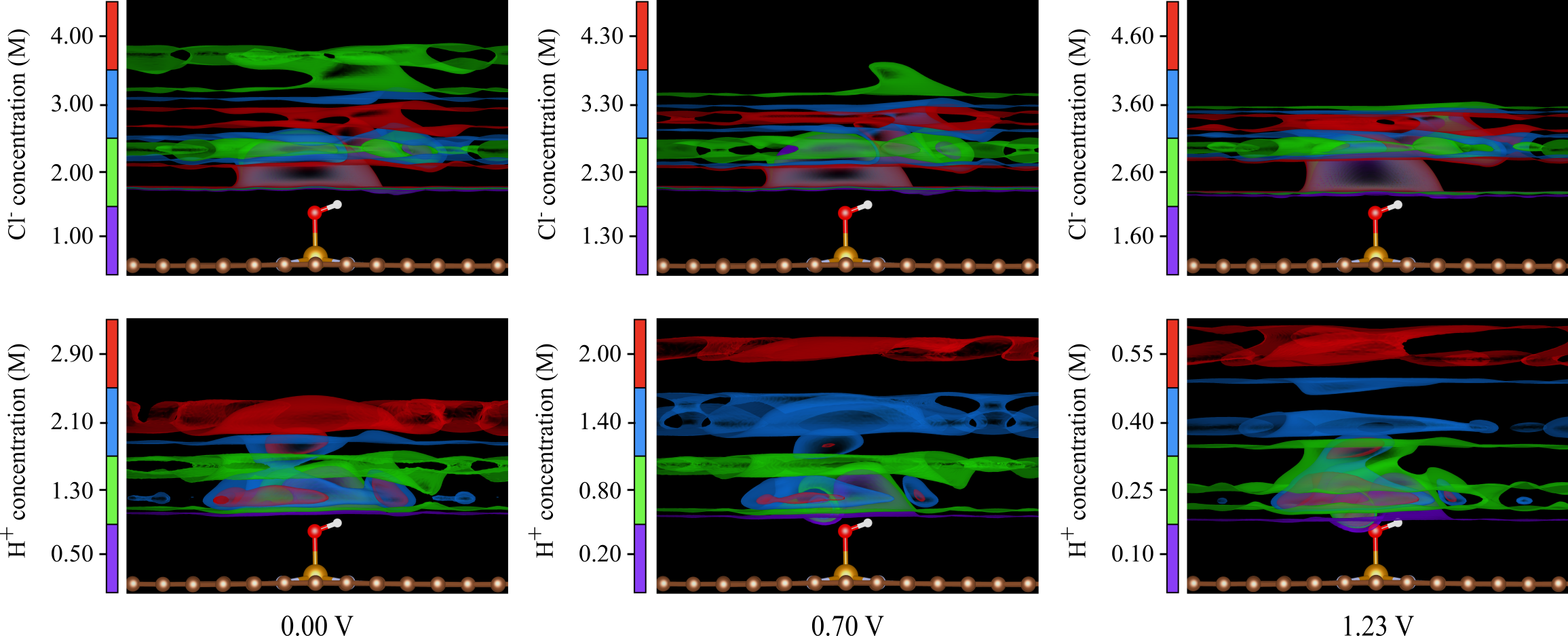}
    \caption{\label{fig:iso} Concentration levels of \ce{H+} (\ce{H3O+}) and \ce{Cl-} ions in 0.5 M HCl at the applied potentials of 0, 0.70, and 1.23 V vs RHE over the Fe$-$\ce{N4}$-$C surface adsorbed with OH. Gold, silver, brown, red, and white spheres represent the Fe, N, C, O, and H atoms, respectively. Note that the range of the concentration is changed depending on the electrolyte ion and applied potential for the visibility.} 
\end{figure}

We begin by describing the EDL formed at the \ce{HCl}/graphene solution interface from our ESM-RISM calculations.
In Figure~\ref{fig:distribution} (a), the distributions of \ce{H3O+} and \ce{Cl-} in the HCl solution on Fe$-$\ce{N4}$-$C with different electrolyte concentrations are depicted.
%
%
We can see that ESM-RISM is capable of determining the detailed statistical description of the EDL structure for a given electrolyte concentration.
We also plot the electrostatic potential of the system in Figure~\ref{fig:distribution} (b), which shows the spatial variation of potential in the EDL depending on the electrolyte concentration, applied potential, and its convergence far from the interface.
%
%
%
To illustrate how the electrolyte ions distribute and respond to the applied potential, we depict the spatial distributions of \ce{Cl-} and \ce{H+}/\ce{H3O+}  ions in the 0.5 M HCl in contact with OH-adsorbed Fe$-$N$_4-$C at different potentials (Figure \ref{fig:iso}).
As anticipated, we can see that as the potential is increased from 0 to 1.23 V vs RHE, the concentration of the \ce{Cl-} ion near the surface increases, while \ce{H+}/\ce{H3O+} is distributed further away.
The electrolyte ions in EDL distribute in a complex way and as a result of the screening, the interfacial electric field (potential) shows nonuniform behavior.
%
%
\subsection{ORR mechanism}
\begin{figure}[h]
   \includegraphics[width=1\columnwidth]{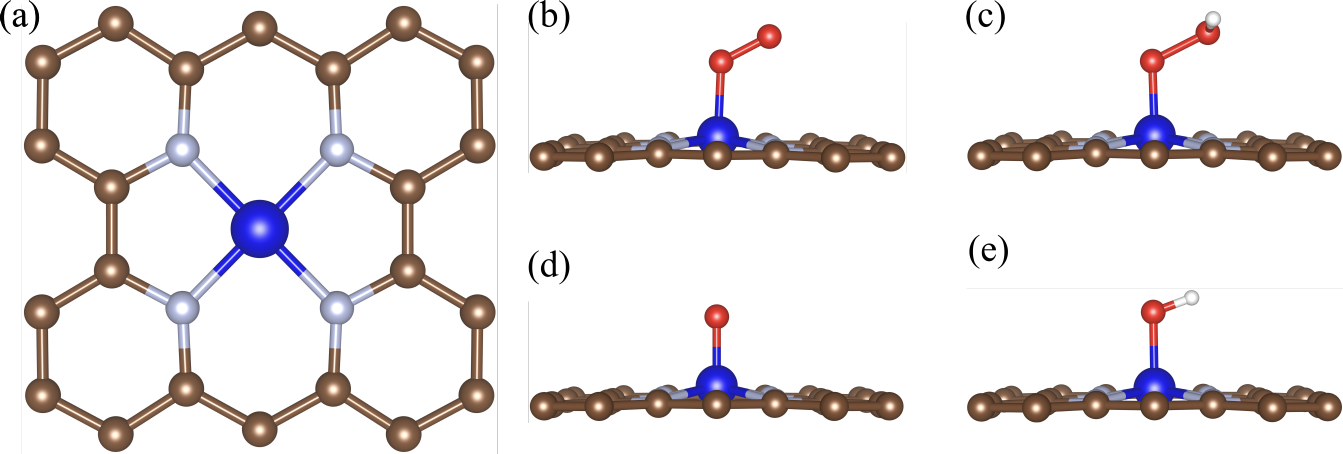}
    \caption{\label{fig:TMN4} Structures of the active site and ORR intermediates. (a) clean TM$-$\ce{N4}$-$C (TM = Fe, Co) active site. (b) $^*$\ce{O2}, (c) $^*$\ce{OOH}, (d) $^*$\ce{O}, and (e) $^*$\ce{OH}. The blue, silver, brown, red, and white atoms represent TM, N, C, O and H, respectively.} 
\end{figure}
We here describe how to derive the free energy of the reaction with ESM-RISM 
(See the Supporting Information for further details).
%
On Fe$-$\ce{N4}$-$C and Co$-$\ce{N4}$-$C catalysts (Figure \ref{fig:TMN4}a), we assume that ORR follows the four-electron associative mechanism and its elementary steps are written as follows:
\begin{align}
        &^{\ast} + \ce{O2(g)} + \ce{H+(aq)} + \ce{e}^{-} \rightarrow                     {^{\ast}}\ce{OOH} & (\Delta G_{1})\\
        &{^{\ast}\ce{OOH}} + \ce{H+(aq)} + \ce{e}^{-} \rightarrow {^{\ast}\ce{O}} + \ce{H2O(l)} & (\Delta G_{2})\\
        &{^{\ast}\ce{O}} + \ce{H+(aq)} + \ce{e}^{-} \rightarrow {^{\ast}\ce{OH}} & (\Delta G_{3})\\
        &{^{\ast}\ce{OH}} + \ce{H+(aq)} + \ce{e}^{-} \rightarrow {^{\ast}} + \ce{H2O(l)} & (\Delta G_{4})
\end{align}
where $^{\ast}$ denotes the adsorption site of the catalyst and $\Delta G_{i}~(i=1,4)$ is the corresponding free energy of the reaction.
In constructing the free energy diagram for ORR, only three different intermediates were considered ($^*$\ce{OOH}, $^*$\ce{O}, and $^*$\ce{OH} (Figure \ref{fig:TMN4}b$-$d)), because the steady-state adsorption of the neutral O$_2$ species does not contribute to the potential dependence\cite{kulkarni2018understanding,karlberg2007estimations}.
%
%
In the constant-$N_{\ce{e}}$ method with ESM-RISM, the free energy of these intermediates is calculated as
\begin{equation}
    G_{X} = A_{\mathrm{tot},X} + E_{\mathrm{ZPE},X} + \int_{0}^{T} dT^{\prime} C_{V,X} - TS_{X}
    \label{eqn:gfree_esm-rism}
\end{equation}
where $A_{\mathrm{tot},X}$, $E_{\mathrm{ZPE},X}$, $\int_{0}^{T}dT^{\prime} C_{V,X}$ and $S_X$ are the Helmholtz free energy, the zero-point energy, the integral of the constant-volume heat capacity ($C_{V,X}$) over temperature ($T$), and the entropy of the intermediate adsorbed species $X$, respectively.
%
The pH effect is automatically included in $A_{\mathrm{tot}}$ within the RISM framework.\cite{haruyama2018electrode}
The free energies of the gas-phase molecules (\ce{H2}, \ce{O2}, and \ce{H2O}) are calculated as $G_{Y}$ $=$ $E_{\mathrm{tot},Y} + E_{\mathrm{ZPE},Y} + \int_{0}^{T} dT^{\prime} C_{p,Y} - TS_Y$, where $E_{\mathrm{tot},Y}$, $\int_{0}^{T} dT^{\prime} C_{p,Y}$, and $S_Y$ are the total energy, the integral is over constant-pressure heat capacity ($C_{p,Y}$), and the entropy of the gas-phase species $Y$.
See Table S2 and S3 for the free energy contributions.
%
In the conventional CHE model with a constant number of electrons (constant-$N_{\mathrm{e}}$), the free energies of the reaction are calculated for the neutral systems, and the effect of the applied potential is introduced as $\Delta G_i(U_{\mathrm{RHE}}) = \Delta G_i + n \vert \ce{e} \vert U_{\mathrm{RHE}}$, where $\Delta G_i$ is the free energy of the reaction for the $i$th step at $\mathrm{pH} = 0$ and $U_{\mathrm{RHE}}=0$, with $U_{\mathrm{RHE}}$ being the electrode potential with reference to the reversible hydrogen electrode (RHE), $n$ is the number of electrons transferred in each elementary step, and \ce{e} is the elementary charge.
As described above, the pH effect is automatically included in the free energy in our constant-$N_{\mathrm{e}}$ calculation with ESM-RISM, and the reaction free energies are calculated by referencing to the standard hydrogen electrode (SHE), and the potential is converted to $U_{\mathrm{RHE}}$ via $U_{\mathrm{RHE}} = U_{\mathrm{SHE}} - k_{\mathrm{B}} T \ln(10) \times \mathrm{pH} / \vert \ce{e}\vert$.
%

\begin{figure}[h]
\includegraphics[width=1\columnwidth]{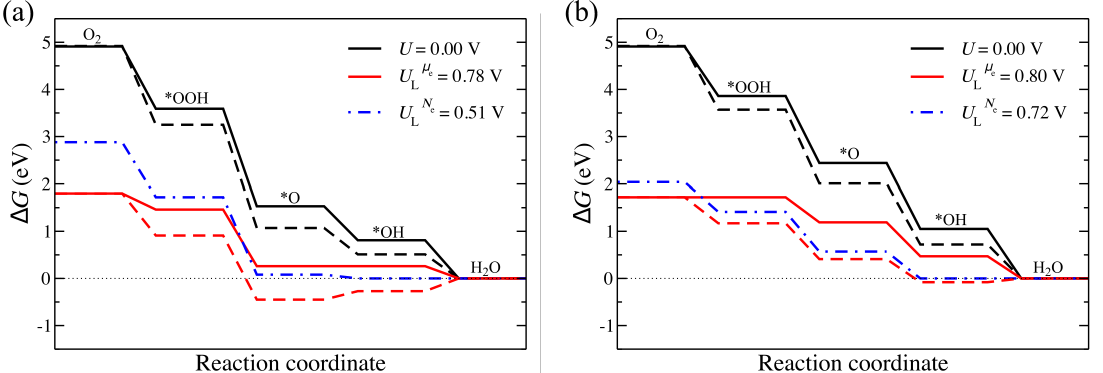}
    \caption{\label{fig:free} Free energy diagram of the ORR for (a) Fe$-$\ce{N4}$-$C and (b) Co$-$\ce{N4}$-$C. The results obtained using the constant-$\mu_{\mathrm{e}}$ and constant-$N_{\mathrm{e}}$ methods are plotted with the solid and dashed lines, respectively. $U_{\mathrm{L}}^{\mu_{\mathrm{e}}}$ refers to the limiting potential estimated using the constant-$\mu_{\mathrm{e}}$ method, while $U_{\mathrm{L}}^{N_{\mathrm{e}}}$ refers to that estimated using the constant-$N_{\mathrm{e}}$ method.} 
\end{figure}

%
In the constant-$\mu_{\mathrm{e}}$ method with ESM-RISM, the electrode potential can be defined by referencing to the inner potential (see Ref.~\cite{haruyama2018electrode} and the Supporting Information) and it is not necessary to use the equilibrium condition $\ce{H+(aq)}+\ce{e}^{-}\leftrightarrow 1/2 \ce{H2(g)}$ at the SHE potential.
In the acidic condition, one can assume that \ce{H+} exists in the form of \ce{H3O+}, and the elementary steps of ORR may be written as
\begin{align}
        &\ce{O2(g)} + \ce{H3O+(aq)} + \ce{e}^{-} \xrightarrow{U}               {^{\ast}\ce{OOH}} + \ce{H2O(l)} & (\Delta G_1(U)) \\
        &{^{\ast}\ce{OOH}} + \ce{H3O+(aq)} + \ce{e}^{-} \xrightarrow{U} {^{\ast}\ce{O}} + 2\ce{H2O(l)} & (\Delta G_2(U))\\
        &{^{\ast}\ce{O}} + \ce{H3O+(aq)} + \ce{e}^{-} \xrightarrow{U} {^{\ast}\ce{OH}} + \ce{H2O(l)} & (\Delta G_3(U))\\
        &{^{\ast}\ce{OH}} + \ce{H3O+(aq)} + \ce{e}^{-} \xrightarrow{U} 2\ce{H2O(l)} & (\Delta G_4(U))
\end{align}
where $U$ on the arrow indicates that the reaction takes place at the constant potential $U$.
%
The free energy of the intermediate species $X$ is calculated as
\begin{equation}
    G_{X}(U) = \Omega_{X}(U) + E_{\mathrm{ZPE},X} + \int_{0}^{T} dT^{\prime} C_{V,X} - TS_X,
    \label{eqn:gfree_esm-risim-c-mu}
\end{equation}
where $\Omega_{X}(U) = A_{\mathrm{tot},X} - \mu_{\ce{e}} \Delta N_{\ce{e}}$ is the electronic grand potential at a constant electron chemical potential $\mu_{\ce{e}}$ with reference to $U$, $A_{\mathrm{tot},X}$ is the Helmholtz free energy, $\Delta N_{\ce{e}}$ is the deviation of the number of electrons from the neutral one,
$E_{\mathrm{ZPE},X}$ is the zero-point energy, $\int_{0}^{T} dT^{\prime} C_{V,X}$ is integral of the constant-volume heat capacity $C_{V,X}$, and $S_X$ is the entropy.
%
See Table S4 for free energy contributions.
%
See also Refs.~\cite{weitzner2020toward,kano2021study,alsunni2021electrocatalytic,brimley2022electrochemical} for grand canonical simulations at the electrochemical interfaces, which inspired this work.
The free energies of the reaction for each elementary step are expressed as
\begin{align}
    & \Delta G_{1}(U) = G_{^{\ast}\ce{OOH}}(U) + G_{\ce{H2O(l)}} - G_{\ce{O2(g)}} - G_{\ce{H3O+(aq)}} - \mu_{\ce{e}} - G_{^{\ast}}(U) \\
    & \Delta G_{2}(U) = G_{^{\ast}\ce{O}}(U) + 2 G_{\ce{H2O(l)}} - G_{^{\ast}\ce{OOH}}(U) - G_{\ce{H3O+(aq)}} - \mu_{\ce{e}} \\
    & \Delta G_{3}(U) = G_{^{\ast}\ce{OH}}(U) + G_{\ce{H2O(l)}} - G_{^{\ast}\ce{O}}(U)  - G_{\ce{H3O+(aq)}} - \mu_{\ce{e}}  \\
    & \Delta G_{4}(U) = G_{^{\ast}}(U) + 2 G_{\ce{H2O(l)}} - G_{^{\ast}\ce{OH}}(U) - G_{\ce{H3O+(aq)}} - \mu_{\ce{e}}
\end{align}
where
\begin{align}
   &\begin{aligned}
         G_{\ce{O2(g)}} = E_{\ce{O2(g)}} + E_{\mathrm{ZPE},\ce{O2(g)}} + \int_{0}^{T} dT^{\prime} C_{P,\ce{O2(g)}} - TS_{\ce{O2(g)}}
   \end{aligned}\\
   &\begin{aligned}
         G_{{^\ast}}(U) =  \Omega_{^{\ast}}(U)
   \end{aligned}\\
   &\begin{aligned}
         G_{^{\ast}\ce{OOH}}(U) =  \Omega_{^{\ast}\ce{OOH}}(U) + E_{\mathrm{ZPE},{^\ast\ce{OOH}}} + \int_{0}^{T} dT^{\prime} C_{V,^{\ast}\ce{OOH}} - TS_{^{\ast}\ce{OOH}}
   \end{aligned}\\
  &\begin{aligned}
         G_{^{\ast}{\ce{O}}}(U) =  \Omega_{^{\ast}\ce{O}}(U) + E_{\mathrm{ZPE},{^\ast\ce{O}}} + \int_{0}^{T} dT^{\prime} C_{V,^{\ast}{\ce{O}}} - TS_{^{\ast}{\ce{O}}}
   \end{aligned}\\
  &\begin{aligned}
         G_{^{\ast}{\ce{OH}}}(U) =  \Omega_{^{\ast}\ce{OH}}(U) + E_{\mathrm{ZPE},{^\ast\ce{OH}}} + \int_{0}^{T} dT^{\prime} C_{V,^{\ast}{\ce{OH}}}  - TS_{^{\ast}{\ce{OH}}}
   \end{aligned}\\
  &\begin{aligned}
         G_{\ce{H2O(l)}} = A_{\ce{H2O(l)}} + E_{\mathrm{ZPE},\ce{H2O(l)}} + \int_{0}^{T} C_{p,\ce{H2O(l)}} \ dT -  TS_{\ce{H2O(l)}} 
   \end{aligned}\\   
   &\begin{aligned}
         G_{\ce{H3O+(aq)}} = A_{\ce{H3O+(aq)}} + E_{\mathrm{ZPE},\ce{H3O+(aq)}}  + \int_{0}^{T} dT^{\prime} C_{p,\ce{H3O+(aq)}} -  TS_{\ce{H3O+(aq)}}
   \end{aligned}
\end{align}
$E_{\ce{O2(g)}}$ is the total energy of \ce{O2} in the gas phase, $\Omega_{^{\ast}}(U)$ is the electronic grand potential of the clean TM$-$\ce{N4}$-$C surfaces ($^{\ast}$), and $\Omega_{^{\ast}\ce{OOH}}(U)$, $\Omega_{^{\ast}\ce{O}}(U)$, and $\Omega_{^{\ast}\ce{OH}}(U)$ are the electronic grand potentials of the TM$-$\ce{N4} surfaces with $^{\ast}$\ce{OOH}, $^{\ast}$\ce{O}, and $^{\ast}$\ce{OH}, respectively.
$A_{\ce{H2O(l)}}$ and $A_{\ce{H3O+(aq)}}$ are the free energies of solvated \ce{H2O} and \ce{H3O+} determined for each electrolyte solution.
%
%
%
%
$E_{\mathrm{ZPE}}$, $\int_{0}^{T} dT^{\prime} C_V$, and $TS$ for the intermediate species  ($^*$\ce{OOH}, $^*$\ce{O}, and $^*$\ce{OH}) were estimated from the harmonic vibrational frequencies calculated using the atomic simulation environment (\textsc{ASE}\cite{larsen2017atomic}) in the vacuum condition, while $E_{\mathrm{ZPE}}$, $\int_{0}^{T} dT^{\prime} C_p$, and $TS$ for \ce{H2O}(l) and \ce{H3O+}(aq) were estimated using ASE in the implicit solution within ESM-RISM and for \ce{O2(g)} and \ce{H2(g)} in the vacuum condition 
($C_p$ is the constant-pressure heat capacity).
%
We note that the entropies of \ce{H2O(l)} and \ce{H3O+(aq)} species were scaled by a factor of 0.65 as discussed in Refs. \cite{garza2019solvation,weitzner2020toward}.
$\mu_{\ce{e}}$ is the target Fermi energy with reference to applied potential $U$:
At $U=$ 0 V vs RHE in 1 M HCl solution (pH $=$ 0), $\mu_{\ce{e}} = -5.33$ eV (see the Supporting Information for details).
The free energy of the reaction for water formation was calculated to be $-$4.91 eV using the current formulation ($\ce{O2(g)} + 4\ce{H3O+(aq)} + 4\ce{e-}  \rightarrow 6\ce{H2O(l)}$) and the constant-$\mu_{\mathrm{e}}$ method, in good agreement with the experimental value of $-$4.92 eV.
Computationally, the thermodynamic limiting-potential (${U_\mathrm{L}}$) is referred to as the highest potential at which all of the reaction steps in the free energy diagram are downhill, and the difference between the ORR equilibrium potential ($U = 1.23$ V vs RHE) and ${U_\mathrm{L}}$ is called the overpotential, calculated by $\eta = 1.23 - {U_\mathrm{L}}$.
In the constant-$N_{\mathrm{e}}$ method, ${U_\mathrm{L}}$ can be simply determined as the minimum free energy change among all of the elementary steps of the ORR.
In the constant-$\mu_{\mathrm{e}}$ method, on the other hand, one has to iterate the calculations with different applied potentials $U$ until all of the elementary steps is exothermic.
Note that, in the constant-$\mu_{\mathrm{e}}$ method, the total charge of the system changes depending on the applied potential, in contrast to the conventional CHE/constant-$N_{\mathrm{e}}$ method, where the calculations are performed in the neutral condition, and the effect of potential is taken into account in an ad hoc manner.
Figure~\ref{fig:free} shows the free energy diagrams of the ORR for TM$-$\ce{N4}$-$C obtained using the constant-$\mu_{\mathrm{e}}$ method as well as the constant-$N_{\mathrm{e}}$ method with ESM-RISM (See also 
Figure S2
).
The potential-determining step (PDS), limiting potential ($U_{\mathrm{L}}$), and overpotential ($\eta$) are summarized in Table \ref{table:overpotential}, along with those obtained in the vacuum condition\cite{abidin2022comparative} for comparison.
We found that the inclusion of solution is crucial:
Indeed, the PDS changes from ${^{\ast}\ce{O}}\rightarrow{^{\ast}\ce{OH}}~(\Delta G_3)$ to ${^{\ast}\ce{OH}}\rightarrow{\ce{H2O}}~(\Delta G_4)$ for Fe$-$\ce{N4}$-$C, and from $\ce{O2}\rightarrow{^{\ast}\ce{OOH}}~(\Delta G_1)$ to ${^{\ast}\ce{OH}}\rightarrow{\ce{H2O}}~(\Delta G_4)$ for Co$-$\ce{N4}$-$C, even without the electrolyte ions (see
Figure S3
and
Table S5, S6
for detailed comparison).
The associated $U_{\mathrm{L}}$ ($\eta$) changes from 0.66 to 0.51 V (0.57 to 0.72 V) and 0.93 to 0.72 V (0.30 to 0.51 V) for Fe$-$\ce{N4}$-$C and Co$-$\ce{N4}$-$C in contact with 0.5 M \ce{HCl} solution, respectively.
We found that the effect of the electrolyte concentration on $U_{\mathrm{L}}$ and $\eta$ is not significant within the current ESM-RISM framework, and the limiting potential is almost unchanged when the concentration is increased (see 
Table S5 and S6
for the results of different electrolyte concentrations).
With the constant-$N_{\mathrm{e}}$ method, the predicted $\eta$ for Co$-$\ce{N4}$-$C (0.51 V) is significantly lower than that for Fe$-$\ce{N4}$-$C (0.72 V) and competitive to that for Pt(111) (0.48 V)\cite{hansen2008surface}, indicating that the former is more reactive than the latter.
Our results are in good agreement with the recent results obtained using DFT with implicit solvation models by Haile et al.\cite{haile2022role} and Patniboon and Hansen\cite{patniboon2021acid}, where $\eta$ for the Co$-$\ce{N4}$-$C is calculated to be 0.57 V and 0.46 V, respectively, and with those by Yang et al.\cite{yang2020unveiling}, Wang et al.\cite{wang2019self}, and Patniboon and Hansen\cite{patniboon2021acid}, where $\eta$ for Fe$-$N${_4}$$-$C is predicted to be as high as 0.70 V, 0.83 V, and 0.83 V, respectively.
%

%
%
With the constant-$\mu_{\ce{e}}$ method, we found that there is a competition between ${^{\ast}\ce{O}}\rightarrow{^{\ast}\ce{OH}}$ and ${^{\ast}\ce{OH}}\rightarrow{\ce{H2O}}$ for the Fe$-$\ce{N4}$-$C, and ${\ce{O_2}}\rightarrow{^{\ast}\ce{OOH}}$ and ${^{\ast}\ce{OH}}\rightarrow{\ce{H2O}}$ for the Co$-$\ce{N4}$-$C.
%
%
In the case of Fe$-$\ce{N4}$-$C, the PDS changes to ${^{\ast}\ce{O}}\rightarrow{^{\ast}\ce{OH}}$ ($\Delta G_3$) with $U_{\mathrm{L}}$ ($\eta$) is predicted to be 0.78 V (0.45 V), while in the case of Co$-$\ce{N4}$-$C, it changes to ${\ce{O_2}}\rightarrow{^{\ast}\ce{OOH}}$ ($\Delta G_1$) with $U_{\mathrm{L}}$ ($\eta$) of 0.80 V (0.43 V).
Given its competing nature, we presume that the actual PDS and associated $U_{\mathrm{L}}$ may change depending on the local environment (such as defects near the active site or substrate below the carbon surface), which is a subject for future investigation.
%
%
%
%
Our results indicate that using the constant-$\mu_{\mathrm{e}}$ method with ESM-RISM, Fe$-$\ce{N4}$-$C and Co$-$\ce{N4}$-$C show comparable activity, in contrast to those obtained using the constant-$N_{\mathrm{e}}$ method.
The highest experimental ORR half-wave potential ($E_{1/2}$) which can be compared with the calculated limiting potential, is reported to be 0.75$-$0.83 (0.82$-$0.92) V vs RHE for Fe$-$N$-$C\cite{zhou2021fe,jiao2021chemical,xie2020performance,yin2020construction,wang2017design} (Co$-$N$-$C\cite{chen2021atomic,yin2020construction,wang2017design}), and our calculated $U_{\mathrm{L}}$'s are in reasonable agreement, considering the uncertainty of the active sites of the catalysts in the experiments.
%
%
%
%
Note that, the $E_{1/2}$ for Co$-$N$-$C is reported to be comparable\cite{wang2017design} with or slightly lower\cite{yin2020construction} than that for Fe$-$N$-$C, when both the catalysts are synthesized and analyzed within the same experimental procedures\cite{yin2020construction,wang2017design}.
Thus, we conclude that the constant-$\mu_{\mathrm{e}}$ method with ESM-RISM gives results which are more consistent with the experiments and improves upon the conventional simulation based on the constant-$N_{\mathrm{e}}$ method.
%
%
%
\begin{table}
  \caption{The potential-determining step (PDS), limiting potential ($U_{\mathrm{L}}$), and overpotential ($\eta$) of ORR estimated for TM$-$\ce{N4}$-$C in vacuum and those for TM$-$\ce{N4}$-$C in contact with 0.5 M \ce{HCl} solution obtained using the constant-$N_{\mathrm{e}}$ and constant-$\mu_{\mathrm{e}}$ methods with ESM-RISM. $^{\mathrm{a}}$The results obtained with the constant-$N_{\mathrm{e}}$ and vacuum condition are taken from Ref.~\cite{abidin2022comparative}.
  \label{table:overpotential}}
  \begin{tabular}{lccc}
    \hline
    & constant-$N_{\mathrm{e}}$ (vacuum)$^{a}$  & constant-$N_{\mathrm{e}}$ (ESM-RISM) &  constant-$\mu_{\mathrm{e}}$ (ESM-RISM)\\
    \hline
    & \multicolumn{3}{c}{Fe$-$\ce{N4}$-$C}\\
    PDS & $^{\ast}\ce{O} \rightarrow  {^{\ast}\ce{OH}}~(\Delta G_3)$ & ${^{\ast}\ce{OH}} \rightarrow \ce{H2O}~(\Delta G_4)$ & $^{\ast}\ce{O} \rightarrow ^{\ast}\ce{OH}~(\Delta G_3)$ \\
    $U_{\mathrm{L}}$  & 0.66 V & 0.51 V & 0.78 V \\
    $\eta$ & 0.57 V & 0.72 V & 0.45 V  \\
    & \multicolumn{3}{c}{Co$-$\ce{N4}$-$C}\\
        PDS & $\ce{O2} \rightarrow {^{\ast}\ce{OOH}}~(\Delta G_1)$ & ${^{\ast}\ce{OH}} \rightarrow \ce{H2O}~(\Delta G_4)$ & $\ce{O2} \rightarrow {^{\ast}\ce{OOH}}~(\Delta G_1)$ \\
    $U_{\mathrm{L}}$ & 0.93 V &  0.72 V & 0.80 V \\
    $\eta$ &  0.30 V & 0.51 V & 0.43 V \\
    \hline
  \end{tabular}
\end{table}

\begin{figure}[ht]
\includegraphics[width=1\columnwidth]{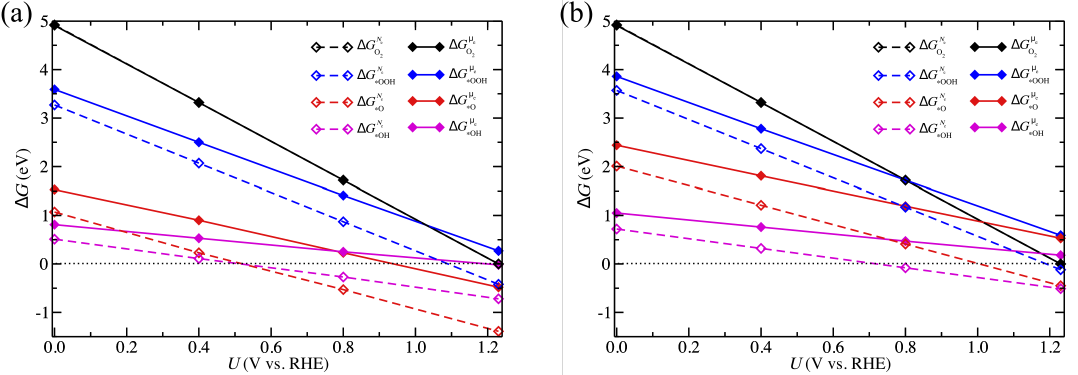}
    \caption{\label{fig:Free-Fe-Co} Free energies of the reaction intermediates relative to the final state ($\Delta G$'s) as functions of applied potentials ($U$) for (a) Fe$-$\ce{N4}$-$C and (b) Co$-$\ce{N4}$-$C at 0.5 M HCl obtained using the constant-$N_{\ce{e}}$ ($\Delta G^{N_{\ce{e}}}$) and constant-$\mu_{\ce{e}}$ ($\Delta G^{\mu_{\ce{e}}}$) methods. The lines are guides for eyes.} 
\end{figure}

To understand the origin of the difference between the results obtained using the constant-$N_{\mathrm{e}}$ and constant-$\mu_{\mathrm{e}}$ methods, we first compare the free energies of ORR intermediates ($\Delta G$'s) with and without solvent 
(Table S5 and S6).
%
By the introduction of water without electrolyte ions in the neutral condition with ESM-RISM, the $\Delta G$'s for OOH, O, and OH decreases for both TM$-$\ce{N4}$-$C's, indicating that water stabilizes all of the adsorbates.
By further introduction of the electrolyte ions with ESM-RISM in neutral conditions, the $\Delta G$'s decrease consistently (at all of the potentials considered in this work), suggesting that the electrolyte ions contribute to the stabilization of the reaction intermediates.
We calculated the charge distribution in the electrolyte solution (Figure S4) and found that the charge of the electrolyte solution distributes according to the charges of the atoms in the molecule and stabilizes the reaction intermediates.
We also calculated $\Delta G$'s 
for the ORR intermediates (Figure~\ref{fig:Free-Fe-Co};Tables S5 and S6)
with the constant-$\mu_{\mathrm{e}}$ and ESM-RISM methods from 0 to 1.23 V vs RHE and found that at 0 V vs RHE, all the adsorbates on TM$-$\ce{N4}$-$C are less stable than those in the neutral state obtained using the constant-$N_{\ce{e}}$ method (larger $\Delta G$'s).
By increasing the electrode potential, $\Delta G$'s decrease as they should.
%
%
With the constant-$N_{\mathrm{e}}$ method (within the CHE framework), $\Delta G$'s and thus free energies of the reaction at an applied potential are solely determined by the adsorption free energy under the neutral condition and the potential dependence of $-n \vert \ce{e} \vert U$ regardless of the actual charge state at the potential $U$.
With the constant-$\mu_{\mathrm{e}}$ method, the potential dependence is almost linear, but the slope deviates from $-n \vert \ce{e} \vert $ depending on the adsorbate, resulting in different limiting potentials/overpotentials from those obtained using the constant-$N_{\mathrm{e}}$ method.
%
To understand the difference between the results obtained using the constant-$N_{\ce{e}}$ and constant-$\mu_{\ce{e}}$ methods, we inspected the total charge and found that the system is positively charged even at $U=0$ V vs RHE and gets more positively charged at $U > 0$ V vs RHE (Figure S5 and S6; Table S7).
See also Figure S7 and S8 for the change in the solvent charge distribution in response to the applied potential.
%
As a result, TM$-$\ce{N4}$-$C adsorbed with the ORR intermediate becomes less and less stable relative to clean TM$-$\ce{N4}$-$C as $U$ is increased (Figure S9), and the slope of the potential dependence of $\Delta G$'s deviates from $-n \vert \ce{e} \vert$.
%


%
We further analyzed the electronic structures at applied electrode potentials by calculating the density of states (DOS) of TM$-$\ce{N4}$-$C with and without reaction intermediates (Figure S10 and S11).
%
%
The DOSs show clear potential dependence, which makes the free energies of the reaction intermediates thereby reaction free energies of the elementary steps of ORR different from those obtained using the constant-$N_{\mathrm{e}}$ method.
\section{Summary}
We present a DFT study on the mechanism of ORR on Fe$-$\ce{N4}$-$C and Co$-$\ce{N4}$-$C in contact with an 
acidic
solution under an applied potential, which has been enabled by ESM-RISM.
%
We found that the solvent effect is essential in modeling the ORR, as PDS changes from ${^{\ast}\ce{O}} \rightarrow {^{\ast}\ce{OH}}$ (${\ce{O2}} \rightarrow {^{\ast}\ce{OOH}}$) to ${^{\ast}\ce{OH}}\rightarrow \ce{H2O(l)}$ (${^{\ast}\ce{OH}}\rightarrow \ce{H2O(l)}$) for Fe$-$\ce{N4}$-$C (Co$-$\ce{N4}$-$C) when the implicit solvent is introduced, but the effect of the electrolyte concentration is minor on the limiting potential.
%
%
We also found that, with the constant-$\mu_{\ce{e}}$ method combined with ESM-RISM, the reaction intermediates become competitive, and the resulting PDSs are ${^{\ast}\ce{O}} \rightarrow {^{\ast}\ce{OH}}$ and ${\ce{O2(g)}\rightarrow^{\ast}\ce{OOH}}$ for Fe$-$\ce{N4}$-$C and Co$-$\ce{N4}$-$C, respectively.
%
%
The calculated limiting potentials are 0.78 and 0.80 V for both Fe$-$\ce{N4}$-$C and Co$-$\ce{N4}$-$C, which are comparable, in contrast to the results obtained with the constant-$N_{\mathrm{e}}$ method and those in the literature, in which superior catalytic activity of Co$-$\ce{N4}$-$C has been predicted.
The origin of the discrepancy between the results obtained with the two methods is mainly the charge state (different from the neutral one) at a fixed electrode potential, which can be determined through the constant-$\mu_{\mathrm{e}}$ method with ESM-RISM.
%
%
This work clarifies the roles of solvent and electrolyte ions within the RISM framework and demonstrates the importance of the variable charge state through the constant potential calculation.
However, studying specific adsorption of electrolyte ion and the effect on the active site\cite{kamat2022acid,holst2018enhanced,patniboon2023effects} is challenging with the plain RISM framework.
Further investigation of the electrolyte effect on the catalytic activity by e.g., explicitly treating anions with ESM-RISM will be of great interest.
Nevertheless, we anticipate that this work is a step toward more realistic simulations of the electrochemical reactions.
\begin{acknowledgement}
This work was partly supported by Grant in Aid for Scientific Research on Innovative Areas "Hydrogenomics" (Grant No.~JP18H05519) and “Program for Promoting Researches on the Supercomputer Fugaku” (Fugaku battery \& Fuel Cell Project) from the Ministry of Education, Culture, Sports, Science, and Technology, Japan (MEXT).
A.F.Z.A. acknowledges the financial support from MEXT.
Calculations were performed using the facilities of the Supercomputer Center, the Institute for Solid State Physics, the University of Tokyo, and of the Cybermedia Center, Osaka University.
\end{acknowledgement}

\section{Data availability}
 The data underlying this study are openly available in Materials Cloud with the identifier: DOI: 10.24435/materialscloud:7j-bt. 

\begin{suppinfo}
Supporting Information is available. 
%
Computational methods including the computational detail, SHE potential, formulation of the free energy of the reaction with constant-$N_{\ce{e}}$ and constant-$\mu_{\ce{e}}$ methods;
Comparison of free energy diagrams obtained using different methods;
Free energy contributions and free energies of the reaction intermediates at different electrode potentials;
Solvent charge distribution, total charge of the system at different electrode potentials, grand potentials as functions of chemical potential, and electronic structures (densities of states) for clean TM$-$\ce{N4}$-$C's, Fe$-$\ce{N4}$-$C adsorbed with $^{\ast}$OH, and Co$-$\ce{N4}$-$C with $^{\ast}\ce{OOH}$.
\end{suppinfo}

\bibliography{achemso-demo}

\end{document}